%% file: qwspectrum.tex
\documentclass[a4paper,12pt]{article}

\usepackage[pdftex]{graphicx}
\usepackage[english]{babel}
\usepackage{amsmath,amssymb} %,amsfonts}
\usepackage[T1]{fontenc}

\usepackage{caption}

\author{Lauri Lehman}
\title{Role of the coin in the spectrum of quantum walks}
\date{}

\newcommand{\bra}[1]{\big\langle#1\big\vert}
\newcommand{\ket}[1]{\big\vert#1\big\rangle}
\newcommand{\tr}{\text{Tr}}

\begin{document}

% em: width of M in the font; pt: 1/72 inch or 1/3 mm
% ex: height of x in the font

\maketitle

\abstract{
The most elementary quantum walk is characterized by a 2-dimensional
unitary coin flip matrix, which can be parameterized by 4 real variables.
The influence of the choice of the coin flip matrix on the time evolution
operator is analysed in a systematic way.
By changing the coin parameters, the dispersion and asymmetry of eigenvalues
of the time evolution operator can be tuned in a controlled way.
The reduced eigenvectors in coin space are distributed along trajectories
on the surface or inside the Bloch sphere, depending on the
degeneracy of the spectrum.
At certain values of the coin parameters the spectrum of the time
evolution operator becomes 2-fold degenerate, but there might
exist unique eigenvalues at the top and bottom of each
quasi-energy band.
The eigenstates corresponding to such eigenvalues are robust
against arbitrary temporal variations in the bias parameter of the
coin, as long as rest of the parameters remain unchanged.
}

\section{Introduction}

Quantum information and computation is a field which uses concepts
from computer science to understand the ability of quantum systems
to store and transmit information and to perform computations.
The quantum walk is a prime example of this vast field:
it was conceived by adopting the concept of random walks from
classical information theory to quantum dynamics.
The quantum mechanical version of the random walks shows interference
effects which might allow faster algorithms than those based on the
classical random walks.

Quantum walks have been studied intensively for over a decade.
The discrete-time quantum walk was introduced as a unitary process
where a single particle, called the walker, hops on a lattice and
the direction of movement is decided by its internal state,
called the coin degree of freedom \cite{abnvw2001,aakv2001,konno2002}.
The propagation of the walker is ballistic as opposed to classical
random walks where the walker propagates diffusively.
The proposition of quantum walks led to studies on whether this
effect could be used to design computational algorithms with a quantum speedup
\cite{skw2003,kendon2006,ambainis2008}.
The simplicity of the quantum walk model has also inspired many studies
on various aspects of quantum dynamics and quantum information,
such as localisation \cite{ikk2004,jm2010},
quantum to classical transition \cite{bca2003-1} and
quantum state transfer \cite{kw2011,bpak2014,yg2014,zqblx2014}.
Moreover, it has been shown that the quantum walk can be used to
perform universal quantum computation \cite{childs2009,lcetk2010}.
Overviews on the properties of quantum walks can be found in
Refs. \cite{konno2008,venegas2012}.
They have also been implemented experimentally with
optical lattices \cite{kfcsamw2009,gaswwma2013},
trapped ions \cite{smsgehs2009,zkgsbr2010} and
optical photons \cite{scpgmajs2010,bflkaw2010,plmmppzliwbsto2010}.

The quantum walks come in many varieties.
Ultimately the walks are defined by the underlying lattice structure
and the transition rules for moving between the lattice sites.
Together they define the time evolution operator, which describes
the evolution of the wave function between subsequent time steps.
The behaviour of the walk is entirely determined by the eigenvalues
and -vectors of the time evolution operator, and the initial
state of the walker.
The purpose of this paper is to explore the variety of the spectrum
of the time evolution operator with arbitrary 2-dimensional unitary coins
for translationally invariant discrete-time quantum walks on the $N$-cycle.
The eigenvalues and eigenvectors for this problem have been solved
by Tregenna et al \cite{tfmk2003}, who also derived formulas for the
asymptotic time-averaged probability distributions in terms of
arbitrary coin parameters.
In this paper, we show that by choosing an appropriate parameterization
for the coin, each parameter plays a distinct role in the properties
of the eigenvalues and eigenstates.
The solution allows therefore to design quantum walks which have precisely
the desired spectral features.
For example, the gaps between the eigenvalues determine important
properties of the walk, such as mixing time \cite{aakv2001}, and it is
desirable to have control over such features.

To further investigate the properties of the eigenstates, we obtain
the Bloch representation of the reduced eigenvectors in coin space,
and study the properties of the eigenvectors as a function of coin
parameters.
The eigenvectors are found to be distributed along well-defined
trajectories on the surface or inside the Bloch ball, depending
on the degeneracy of the spectrum.
With special attention to degenerate eigenvalues, we point out that
there may also exist unique eigenvalues in an otherwise degenerate
spectrum.
The eigenstates corresponding to the unique eigenvalues have very
special properties, since they remain fixed for arbitrary variations
of the bias parameter of the coin.
These states are thus protected from arbitrary temporal variations
of the bias parameter, if the rest of the parameters remain fixed.

This article is organized as follows.
The discrete-time quantum walk and the parameterization of the coin
is introduced in Sec. \ref{sec:malli}.
The eigenvalue spectrum is studied in Sec. \ref{sec:ominaisarvot},
and the Bloch representation of the eigenvectors is given
in Sec. \ref{sec:ominaisvektorit}.
The protected eigenstates are discussed in Sec. \ref{sec:erikoistilat},
and the conclusions are presented in Sec. \ref{sec:johtopaatokset}.

\section{Discrete-time quantum walk} \label{sec:malli}

The discrete-time quantum walk proceeds by repeated steps of coin
flipping and translation to the left or right, depending on the state
of the coin \cite{abnvw2001,aakv2001}.
The walk is assumed to take place on a 1-dimensional lattice with
periodic boundary conditions, and translation invariance is
imposed by taking the same coin at every site of the lattice.
The wave function of the walker can be written as
$\ket{\psi(t)} = \sum_{x,c} \psi_{xc}(t) \ket{x}\ket{c}$,
where the amplitudes $\psi_{xc}(t)$ are complex numbers.
The number of sites is $N$ and the coin variable $c$ can take
values $ \{ 0, 1 \}$.
The total Hilbert space is a tensor product of the spatial and
coin spaces, $\ket{\psi(t)} \in \mathbb{H}_X \otimes \mathbb{H}_C$,
with dimension $2N$.

The time-independent step operator $U$ evolves the wave function
by one time step: $\ket{\psi(t+1)} = U \ket{\psi(t)}$.
It can be written as $U = \mathcal{T F}$, where the coin flip
acts on the coin degrees of freedom only: $\mathcal{F} = I \otimes F$.
The coin flip $F$ is represented by a 2-dimensional unitary matrix.
Such a matrix has generally 8 real components, since each of the 4
complex elements can be represented by 2 real numbers.
Requiring the matrix to be unitary, $FF^\dag = F^\dag F = I$,
imposes 4 independent conditions for the elements,
and therefore the matrix is represented by 4 independent
real parameters.
One of the parameters is just the total complex phase of the
matrix, and it will be ignored since it does not have
any effect on the time evolution of the probability
distribution of the walker.
The rest of the parameters can be chosen in infinitely many ways,
but as will be seen below, the following parameterization is
particularly insightful when studying the spectrum of $U$:
\begin{equation} \label{eq:kolikko}
F\; =\;
e^{i \beta} \begin{pmatrix}
\sqrt{R}\; e^{i\alpha} & \sqrt{1-R}\; e^{-i\beta} \\
-\sqrt{1-R}\; e^{i\beta} & \sqrt{R}\; e^{-i\alpha}
\end{pmatrix}
\end{equation}
The range of the parameter $R$ is $[0,1]$ and it is known as the
bias parameter, as it sets the weight between
the coin components 0 and 1.
The angular variables are on the interval $[0,2\pi)$.
The coin flip matrix introduces mixing between the coin components
$\ket{0}$ and $\ket{1}$, which correspond to left and right
directed motion of the walker, respectively.

The conditional shift operator $\mathcal{T}$ moves the walker
to the left or right depending on its coin state.
It can be written concisely as
\begin{equation}
\mathcal{T} = T_- \otimes P_0 + T_+ \otimes P_1
\end{equation}
where the operator $P_c = \ket{c}\bra{c}$ is a projector to the
coin state $c$, and
$T_{\pm} = \sum_{x=0}^{N-1}$ $\ket{x\pm1} \bra{x}$ is the
translation operator.
Periodic boundary conditions are imposed by defining the
translations at the righthand boundary as
$\ket{N}\bra{N-1} = \ket{0}\bra{N-1}$ and at the lefthand
boundary as $\ket{-1}\bra{0} = \ket{N-1}\bra{0}$.

\section{Eigenvalue spectrum} \label{sec:ominaisarvot}

The eigenvalues of the time evolution operator $U$ can be solved by
standard methods of quantum mechanics.
Translation invariance implies that the wavenumbers $k$ in Fourier
space are good quantum numbers, which means that the eigenvalue
equation for $U$ becomes $2\times2$ block-diagonal, and each block can be
solved independently.
Since $U$ is unitary, its eigenvalues are complex numbers with absolute
value 1 and can be written as $e^{i\lambda}$.
The phases $\lambda$ can be interpreted as quasi-energies, if the
time evolution of the wave function is thought to be generated
by a time-independent Hamiltonian $H$.
The time evolution operator is then given by $U=e^{-iH\Delta t}$,
where the time step is chosen as $\Delta t=1$.
Indeed, the quasi-energies $\lambda$ are distributed
along single-particle energy bands that become continuous as
the size of the lattice approaches infinity.
The eigenvalues and eigenvectors for the $N$-cycle with a similar
parameterization for the coin have been previously solved by Tregenna
et al \cite{tfmk2003}, but the solution is summarized here since
the notation will be useful later.

The eigenvalue problem for the time evolution operator is given by
the equation
\begin{equation} \label{eq:oarvehto0}
U \ket{\psi_\lambda} = e^{i\lambda} \ket{\psi_\lambda},
\end{equation}
where the eigenvectors are denoted by
$\ket{\psi_\lambda} = \sum_{x,c} \psi_\lambda(x,c) \ket{x}\ket{c}
= \sum_{k,c} \widetilde{\psi}_\lambda(k,c) \ket{k}\ket{c}$
and the Fourier-transformed amplitudes are
$\widetilde{\psi}_\lambda(k,c) = \frac{1}{\sqrt{N}}
\sum_{x=0}^{N-1} e^{i2\pi kx/N} \psi_\lambda(x,c)$.
The task is to find the eigenvalue phases $\lambda\in [-\pi,\pi)$
and the components of the eigenvectors $\widetilde{\psi}_\lambda$.
In the Fourier basis, the translation operators $T_-$ and $T_+$ act
diagonally and the wave function is just multiplied by a phase
$e^{-i2\pi k/N}$ and $e^{i2\pi k/N}$, respectively.
Thus, the $2N$ equations in Eq. \eqref{eq:oarvehto0} factorize into
$N$ pairs of decoupled equations, involving only the components
$\widetilde{\psi}_\lambda(k,0)$ and $\widetilde{\psi}_\lambda(k,1)$
at each wavenumber $k$:
\begin{equation} \label{eq:oarvehto}
\begin{pmatrix} g_{00} & g_{01}\\ g_{10} & g_{11} \end{pmatrix}
\begin{pmatrix} \widetilde{\psi}_\lambda(k,0) \\
\widetilde{\psi}_\lambda(k,1) \end{pmatrix} = 0
\end{equation}
where the matrix elements are given by
\begin{align}
g_{00}(\lambda,k)\; =\; &
  \sqrt{R}\; e^{i(\alpha+\beta-2\pi\frac{k}{N})}\;
  -\; e^{i\lambda} \label{eq:g00} \\
g_{10}(k)\; =\; & -\sqrt{1-R}\;
  e^{i(2\beta+2\pi\frac{k}{N})}
  \label{eq:g10} \\
g_{01}(k)\; =\; & \sqrt{1-R}\;
  e^{-i2\pi\frac{k}{N}}
  \label{eq:g01} \\
g_{11}(\lambda,k)\; =\; & \sqrt{R}\;
  e^{-i(\alpha-\beta-2\pi\frac{k}{N})}\; -\;
  e^{i\lambda} \label{eq:g11}
\end{align}
for fixed coin parameters $R,\alpha,\beta$.
Denoting the matrix in  Eq. \eqref{eq:oarvehto} as $G$,
this equation has non-trivial solutions if and only if the
determinant of $G$ is zero.
For most choices of the coin parameters, non-trivial solutions
exist for only one wavenumber $k$, and the eigenvectors are
strictly localized at each $k$.
There exist however special points in the parameter space where
the spectrum becomes degenerate and Eq. \eqref{eq:oarvehto}
allows solutions for two different wavenumbers, as discussed
below.

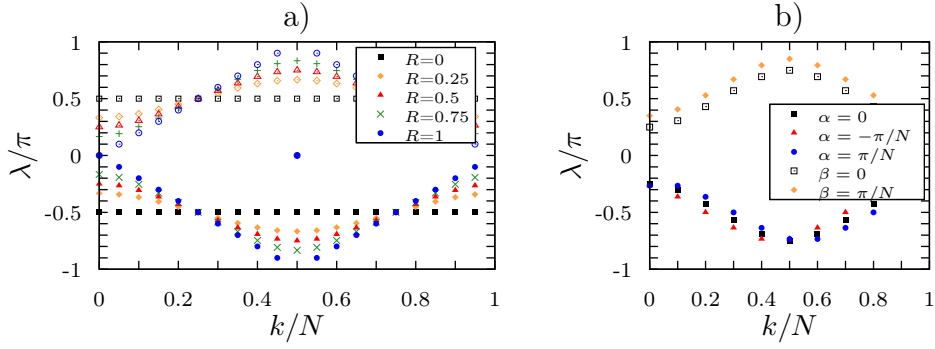
\begin{figure}
\begin{center}
\setlength{\unitlength}{.25\textwidth}
\input{./ominaisarvot_r.tex}
\hspace{2ex}
\input{./ominaisarvot_alphabeta.tex}
\end{center}
\caption{\small Distribution of the eigenvalue phases
$\lambda(k,z)$ as a function $k$ and $z$.
a) Deformation of the spectrum as $R$ varies from 0 to 1.
The eigenvalues with the same $R$ are marked with the same colour;
the negative (positive) values correspond to $z$=1 ($z$=2).
The system size is $N$=20 and the values of the other
coin parameters are $\alpha$=0, $\beta$=0.
b) Dependence on the coin parameters $\alpha$ and $\beta$.
The system size is $N$=10 and $R$=0.5.
For the lower bands $z$=1, $\beta$=0, and for the
upper bands $z$=2, $\alpha$=0.}
\label{fig:ominaisarvot}
\end{figure}

The eigenvalues $e^{i\lambda}$ can be solved from
Eq. \eqref{eq:oarvehto} for each $k$, and there exist two
distinct solutions labeled by an integer $z = \{1,2\}$:
\begin{equation} \label{eq:ominaisarvot}
e^{i\lambda(k,z)}\; =\; e^{i\beta}\Big[ \sqrt{R}\cos(\alpha-2\pi k/N)
+ i\, (-1)^z \sqrt{1-R\cos^2(\alpha-2\pi k/N)} \Big].
\end{equation}
where the term in the square brackets is a complex number
with modulus 1.
The $2N$ eigenvalues are thus labeled by the wavenumber
$k = \{ 0,1, \ldots, N-1 \}$ and the integer $z$.
The eigenvalue phases $\lambda(k,z)$ are plotted as functions of
$k$ and $z$ in Fig. \ref{fig:ominaisarvot} for different values
of the coin parameters.
The eigenvalues for different $k$ are distributed on two
well-defined quasi-energy bands, labeled by the integer $z$.
The shape of the bands is entirely determined by the
coin variables $R$, $\alpha$ and $\beta$, and changing the
total number of sites changes only the relative distances of
the eigenvalues along the band.
As $N\rightarrow\infty$, the eigenvalues on each band become
infinitesimally close to each other.

Each of the parameters $R$, $\alpha$ and $\beta$ correspond to
different properties of the spectrum.
As seen in Fig. \ref{fig:ominaisarvot}a), the parameter $R$ determines
the general shape of the spectrum, interpolating between a completely
flat spectrum at $R = 0$ and a linearly spaced spectrum at $R = 1$.
In the former case the eigenvalues are $N$-fold degenerate, while
in the latter case the gaps between the eigenvalues are degenerate.
In the special case $R = 0$, Eq. \eqref{eq:ominaisarvot} implies
that the phases of the eigenvalues are given by
\begin{equation} \label{eq:oarvvaihe0}
\lambda(k,z)\; =\; \beta + (-1)^z\, \frac{\pi}{2}\quad
\forall k;\quad\quad R = 0,
\end{equation}
and for $R = 1$ they become
\begin{equation} \label{eq:oarvvaihe1}
\lambda(k,z)\; =\;
\begin{cases}
\beta + (-1)^z(\alpha - 2\pi k/N) &
\text{if } \sin(\alpha - 2\pi k/N) \geq 0 \\
\beta - (-1)^z(\alpha - 2\pi k/N) &
\text{if } \sin(\alpha - 2\pi k/N) < 0
\end{cases}
\quad\quad R = 1.
\end{equation}
Eq. \eqref{eq:oarvvaihe1} shows clearly that the eigenvalue phases
for $R = 1$ are distributed symmetrically around the value $\beta$,
with equally spaced intervals $\Delta\lambda = 2\pi/N$.

%TODO band dispersion and coefficient of second moment.
% calculate moments (Laplace transform?); numerics

The role of the parameter $\beta$ is just to set the
``base point" of the spectrum, as can be seen directly
from Eq. \eqref{eq:ominaisarvot}.
The effect of changing $\beta$ is that each eigenvalue phase is
shifted by a constant amount $\Delta\lambda = \Delta\beta$, as
illustrated in Fig. \ref{fig:ominaisarvot}b) for the $z=2$ eigenvalues.
The role of the parameter $\alpha$ is slightly more subtle.
As shown in Fig. \ref{fig:ominaisarvot}b) for the $z=1$ eigenvalues,
increasing the value of $\alpha$ increases $\lambda$ for one half
of the spectrum, and decreases $\lambda$ for the other half.
Changing the value of $\alpha$ introduces asymmetry between the
left and right parts of the spectrum.
If the change in $\alpha$ is an integer multiple of $2\pi/N$,
this corresponds to a mere relabeling of the wavenumbers $k$,
which is evident from Eq. \eqref{eq:ominaisarvot}.
Therefore the spectra corresponding to $\alpha=-\pi/N$ and
$\alpha=\pi/N$ in Fig. \ref{fig:ominaisarvot}b) are
equal, except that the values are shifted horizontally by $\Delta k=1$.

For most values of $\alpha$, the eigenvalues on the left and
right side are not equal, and the spectrum is nondegenerate.
Thus, the left-moving and right-moving components of the
walker are distinguishable, and the walk is called \emph{chiral}.

\section{Bloch representation for eigenvectors} \label{sec:ominaisvektorit}

In the previous section, it was seen that each coin parameter
$R,\alpha,\beta$ has a distinct role in the eigenvalue spectrum
of $U$.
It turns out that the coin parameters also control different
features of the eigenstates.
We obtain a geometric picture of the eigenstates which proves
to be very useful in the analysis of the eigenstates.

The eigenstates take different forms in the degenerate
and non-degenerate cases.
The condition that two eigenvalues are equal,
$e^{i\lambda(k,z)}=e^{i\lambda(k'z')}$ leads to the following relations:
\begin{align}
\cos(\alpha-2\pi k/N)\; =\; &
\cos(\alpha-2\pi k'/N) \label{eq:deg-ehto1} \\
(-1)^z \, \sqrt{1-R\cos^2(\alpha-2\pi k/N)}\; =\; &
(-1)^{z'} \, \sqrt{1-R\cos^2(\alpha-2\pi k'/N)}
\label{eq:deg-ehto2}
\end{align}
The second condition is just equal to $z=z'$, since the
signs must be equal on both sides.
The first condition allows two solutions for the wavenumbers,
either $k'=k$ or
\begin{equation} \label{eq:kconj}
k' = -k + \alpha\frac{N}{\pi} \; \mod N.
\end{equation}
Since the wavenumbers are labeled by integers, the above
equation only makes sense if $\alpha$ is an integer multiple
of $\pi/N$.
Therefore the spectrum becomes degenerate at $2N$ discrete
values of $\alpha$:
\begin{equation} \label{eq:alpha}
\alpha = n \, \frac{\pi}{N} \quad\quad
n \in \{ 0, 1, \ldots, 2N-1 \}.
\end{equation}
A similar result has also been obtained by Tregenna et al \cite{tfmk2003}.
In the maximally degenerate case, there are $N$ distinct
eigenvalues, which are labeled by $N/2$ pairs of conjugate
wavenumbers $(k,k')$ and the integer $z$, and the eigenvalues
are labelled by $\lambda(k,k',z)$.
%TODO uniform vs non-uniform asymptotic distribution, localization \cite{ikk2004,ik2005}

Recall from Eq. \eqref{eq:oarvehto} that the components
of the eigenvectors for wavenumber $k$ can be non-zero
iff $\det(G) = g_{00}g_{11} - g_{10}g_{01} = 0$.
By a straightforward calculation, it can be shown that this
condition is equal to Eq. \eqref{eq:deg-ehto1}.
Therefore, if the spectrum is non-degenerate, only the
$k'=k$ component of the eigenvectors is non-zero,
and the eigenvectors are localized at wavenumber for
each eigenvalue $\lambda(k,z)$.
If $\alpha$ is an integer multiple of $\pi/N$, the
$k'$-component of the eigenvectors is also non-zero,
where the conjugate component $k'$ is given by Eq. \eqref{eq:kconj}.

The relations between the two coin components are trivially solved
from the first row of Eq. \eqref{eq:oarvehto}:
\begin{align}
\widetilde{\psi}_{\lambda}(k,1)\; =\; &
-\frac{g_{00}(\lambda,k)}{g_{01}(k)} \,
\widetilde{\psi}_{\lambda}(k,0) \label{eq:psi1} \\
\widetilde{\psi}_{\lambda}(k',1)\; =\; &
-\frac{g_{00}(\lambda,k')}{g_{01}(k')} \,
\widetilde{\psi}_{\lambda}(k',0) \label{eq:psi2} \\
\widetilde{\psi}_{\lambda}(k'',c)\; =\; & 0
\quad\quad\quad\quad\quad \forall \, c, \; k''\neq k,k'. \label{eq:psi3}
\end{align}
where $|g_{01}|>0$ if $R<1$ and $\lambda=\lambda(k,k',z)$.
The condition \eqref{eq:psi2} refers to the non-degenerate case only.
The second row of Eq. \eqref{eq:oarvehto} is then always
true if $\det G = 0$.

To gain some insight to the properties of the eigenstates,
it is useful to move to a geometric picture.
The reduced density matrix of the coin,
$\rho_C = \tr_X\big( \ket{\psi_\lambda} \bra{\psi_\lambda} \big)$,
is a $2\times2$ matrix that can be represented by a vector
$\vec{r}$, pointing to a point inside the Bloch ball \cite{nielsenchuang}.
The density matrix is given by
$\rho_C = \frac{1}{2} \big( I+\vec{r}\cdot \vec{\sigma} \big)$,
where the $\sigma_i$ are the Pauli matrices.
If the spectrum of $U$ is non-degenerate, the eigenstates are
product states of wavenumbers and coin states, the reduced
coin eigenvectors are pure states, and the Bloch vector points
always to the surface of the sphere.
In the degenerate case, the eigenvectors are not necessarily
product states, and the Bloch vector might be inside the sphere.
To obtain an explicit representation for the eigenvectors,
it is necessary to obtain an orthonormal eigenbasis.

From Eqs. (\ref{eq:psi1}--\ref{eq:psi3}), the eigenvectors for
degenerate eigenvalues can be constructed as
\begin{equation} \label{eq:ominaisvektorit}
\ket{\psi_\lambda} \; = \; s \ket{k}
\Big[ \ket{0} - \frac{g_{00}(\lambda,k)}{g_{01}(k)} \ket{1} \Big]
+ e^{i \omega} s' \ket{k'}
\Big[ \ket{0} - \frac{g_{00}(\lambda,k')}{g_{01}(k')} \ket{1} \Big]
\end{equation}
where $s$ and $s'$ are real numbers and the complex phase of the
$\ket{k}\ket{0}$-component is set to 0.
The variables $s$, $s'$ and $\omega$ are free parameters,
and any choice of these parameters gives an eigenvector of $U$,
but $s$ and $s'$ are not independent once the vectors are normalized.
The normalization $\big\langle \psi_\lambda \ket{\psi_\lambda} = 1$
imposes the constraints
\begin{equation}
s' = \sqrt{
\frac{ |g_{01}(k')|^2 -
\big[ |g_{00}(k)|^2 + |g_{01}(k)|^2 \big] s^2}
{|g_{00}(k')|^2 + |g_{01}(k')|^2} }
\end{equation}
\begin{equation} \label{eq:normiehto-r}
0 < s < s_{\text{max}} \quad\quad\quad
s_{\text{max}} = \frac{ |g_{01}(k')| }{ \sqrt{|g_{00}(\lambda,k)|^2+|g_{01}(k)|^2} }
\end{equation}
where the limit $s\rightarrow s_{\text{max}}$ corresponds to $s'\rightarrow0$.
Any vector satisfying these relations is a normalized eigenvector of $U$,
but two eigenvectors with parameters
$\{s_1,s_1',\omega_1\}$ and $\{s_2,s_2',\omega_2\}$ are not necessarily
orthogonal to each other.
But it is always possible to pick two orthogonal eigenvectors
for each doubly degenerate eigenvalue, if these parameters satisfy
\begin{align}
s_2 \; = \; & \sqrt{ \frac{|g_{01}(k)|^2}
{|g_{00}(k)|^2 + |g_{01}(k)|^2} - s_1^2} \\
\omega_2 \; = \; & \omega_1 + \pi.
\end{align}
Any choice of $s_1$ and $\omega_1$ gives then an orthonormal
pair of eigenvectors.
Note that the weights $s_1$ can be different for each eigenvalue, 
as long as they fall on the interval given in Eq. \eqref{eq:normiehto-r}.

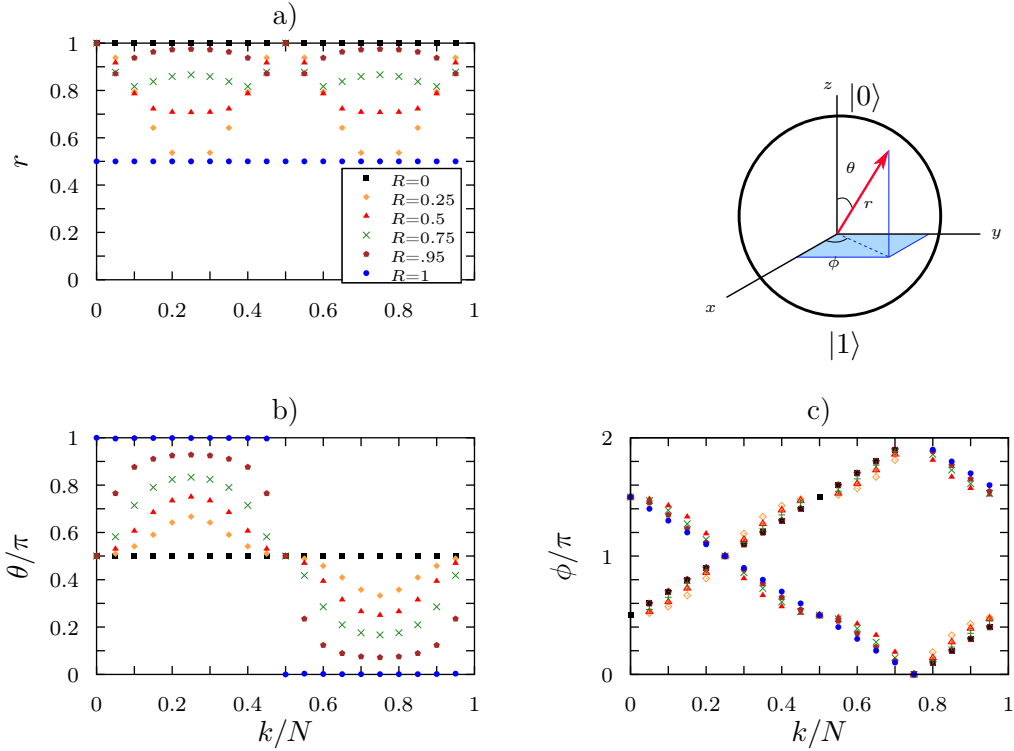
\begin{figure}
\begin{center}
\begin{tabular}{cc}
\input{./ominvekt_r.tex} &
\hspace{4em}
\setlength{\unitlength}{.4\textwidth}
\input{./blochpallo.tex}\\
\input{./ominvekt_theta.tex} &
\input{./ominvekt_phi.tex}
\end{tabular}
\end{center}
\caption{\small Distribution of the reduced coin eigenvectors
on the Bloch sphere for different values of the coin parameter $R$.
The labeling of the markers is the same in all graphs.
In graph c), the filled points correspond to $z$=1 and the
unfilled points correspond to $z$=2.
The weights between the conjugate components are set equal,
$s_i'=s_i$, so that there is no difference between the $z$
components in a) and b).
The system size is $N$=20 and $\alpha$=0, $\beta$=0.}
\label{fig:ominaisvektorit}
\end{figure}

The Bloch sphere representation of the reduced coin eigenvectors can
now be constructed from Eq. \eqref{eq:ominaisvektorit} with the result
\begin{align}
r_x \; = \; & -2\, \text{Re}(\Theta) \\
r_y \; = \; & 2\, \text{Im}(\Theta) \\
r_z \; = \; & s^2\Big( 1 - \Big|\frac{g_{00}(\lambda,k)}{g_{01}(k)}\Big|^2 \Big)
+s'^2\Big( 1 - \Big|\frac{g_{00}(\lambda,k')}{g_{01}(k')}\Big|^2 \Big)
\end{align}
where the function $\Theta$ is defined as
\begin{equation}
\Theta = s^2\frac{g_{00}(\lambda,k)}{g_{01}(k)}
+ s'^2\frac{g_{00}(\lambda,k')}{g_{01}(k')}
\end{equation}
The Bloch ball representation of the eigenvectors is plotted for
different values of the coin parameters in Fig. \ref{fig:ominaisvektorit},
where we have moved to the spherical coordinates
$(r_x,r_y,r_z)\rightarrow(r,\theta,\phi)$.
The eigenvectors are distributed along well-defined trajectories,
which become thicker as the number of lattice points increases,
in analogy to the eigenvalue bands discussed in the previous section.

%TODO k=0,-N/2 ominaistilat kun R=1. eivat havia kun N->aareton?
%TODO special points: $R = 1, g_{00} = g_{11} = 0$. coin components are arbitrary.

It can be seen from Fig. \ref{fig:ominaisvektorit}b) that for small
values of $R$, the coin eigenstates are close to the equator
where $\theta=\pi/2$, and increasing $R$ brings them closer to the
poles of the Bloch ball.
Fig. \ref{fig:ominaisvektorit}c) shows that the eigenstates with the
same $k$ but different $z$ are at approximately same distance from
the point $\phi=\pi$.
The $z$ components of the eigenvectors traverse around the axis in opposite directions,
with the angle $\phi$ increasing or decreasing depending on $z$.
This is in contrast to the non-degenerate case, where the
angle $\phi$ can be shown to be given by
$\phi = \beta - \frac{\pi}{2} + 2\pi\frac{k}{N}$,
ie. independent of $z$.
Note that the eigenvalues corresponding to $k=N/2\pm\delta k$
are equal (see Fig. \ref{fig:ominaisarvot}), and the eigenvectors
for the same value are distributed at equal distance from
the equator.
These considerations hold when the weights $s$ and $s'$ are set equal in Eq.
\eqref{eq:ominaisvektorit}, and it should be noted that changing the parameter $s$
can also deform the trajectories, but these deformations are
not explored here.
%TODO Since any choice of $s$ and $\omega$ gives a valid eigenvector of $U$, these
% parameters can be considered as ``gauge parameters'' which
% have no effect on the observable properties of the system.
%TODO alpha, beta

The degeneracy of the spectrum of $U$ is a signal of some kind of
redundancy in its representation.
This redundancy implies that there exists a non-trivial symmetry
operator $S$ that leaves $U$ unchanged:
\begin{equation} \label{eq:symmetriaop}
SUS^\dag \; = \; U.
\end{equation}
With the solutions for the eigenvectors of $U$, such a symmetry operator can be
constructed by considering the orthogonal eigenvectors for each degenerate
eigenvalue $\lambda(k,k',z)$.
Since the operator $U$ is ``blind'' to the transitions between degenerate
eigenstates $\ket{\psi_\lambda^1}$ and $\ket{\psi_\lambda^2}$,
a non-trivial unitary operator that satisfies Eq. \eqref{eq:symmetriaop} is given by
\begin{equation}
S \; = \; \sum\limits_{\lambda \in \Lambda_n}
\ket{\psi_\lambda}\bra{\psi_\lambda}
+ \sum\limits_{\lambda \in \Lambda_d}
\Big( \ket{\psi_\lambda^2}\bra{\psi_\lambda^1}
+ \ket{\psi_\lambda^1}\bra{\psi_\lambda^2} \Big)
\end{equation}
where $\Lambda_n$ is the set of non-degenerate eigenvalues (if any)
and $\Lambda_d$ is the set of all degenerate eigenvalues.
When written out explicitly, the operator $S$ consists of transitions
between the left-moving and right-moving components $k$ and $k'$,
and simultaneous rotations of the coin.
It must therefore be understood as a joint symmetry which acts
simultaneously on the position and coin degrees of freedom.

\section{Protected eigenstates} \label{sec:erikoistilat}

Even if the rest of the spectrum is degenerate, in some cases the
highest or the lowest eigenvalue might be unique, as seen for
$k=0$ and $k=N/2$ in Fig. \ref{fig:ominaisarvot}a).
Based on the discussion in the previous section, the eigenstates
for unique and degenerate eigenvalues are characteristically
different, since the latter are superpositions of two
wavenumbers $k$ and $k'$.
As seen in Fig. \ref{fig:ominaisvektorit}, the eigenvectors for
the unique eigenvalues correspond to nodes of the trajectories
in the Bloch ball, and these eigenvectors remain fixed as the
value of $R$ changes.

The robustness of the special eigenvectors can be understood
using the results from the previous section.
According to Eqs. (\ref{eq:kconj}--\ref{eq:alpha}), the wavenumbers
for degenerate eigenvalues are arranged into pairs
$(m,n-m)$ where $m=\{0,1,\ldots,M\}$, and
$(n+l,N-l)$ where $l=\{1,2,\ldots,L\}$.
If $n$ is odd, $M=(n-1)/2$ and every wavenumber in the first group
has a matching pair.
But if $n$ is even, $M=n/2-1$ and the wavenumber $k=n/2$ does not
have a pair, and the corresponding eigenvalue is unique.
Similarly if $N-n$ is even, $L=(N-n)/2-1$ and the eigenvalue
corresponding to $k=(N-n)/2$ (modulo $N$) is unique.
The values of $k$ with a unique eigenvalue are summarized
in the following table:
\begin{center}
\begin{tabular}{|r||c|c|}
\hline
 & $N$ even & $N$ odd \\ \hline\hline
$n$ even & $k=\big\{n/2,(N-n)/2\big\}$ & $k=n/2$\\ \hline
$n$ odd & -- & $k=(N-n)/2$ \\ \hline
\end{tabular}
\end{center}
For example in Fig. \ref{fig:ominaisarvot}a), $n=0$ and $N=20$,
and the eigenvalues at $k=0$ and $k=10$ are unique.

The widely used Hadamard coin is a curious special case in the
parameter space.
In the parameterization introduced in Eq. \eqref{eq:kolikko},
it corresponds to the values
$\{R,\alpha,\beta\} = \{1/2, 3\pi/2, \pi/2\}$.
If $N$ is an integer multiple of 4, this choice implies that the spectrum
is degenerate and there are 2 unique eigenvalues in the spectrum,
since the value $n=3N/2$ is an even integer, see Eq. \eqref{eq:alpha}.
If $N$ is an odd integer multiple of 2, the spectrum is fully degenerate
since $n$ is an odd number.
For odd $N$, the parameter $\alpha$ is not an integer multiple of $\pi/N$,
and the spectrum is always non-degenerate.

The robustness of the eigenstates is explained by the behaviour of
the matrix elements of $G$.
Recall from last section that the components of the eigenstates for
unique eigenvalues must satisfy Eq. \eqref{eq:psi1}.
If $\alpha=n\pi/N$ and $k=n/2$, then $\alpha-2\pi k/N=0$ and
it is easy to show that $g_{00}/g_{01}=(-1)^ze^{i(\alpha+\beta+3\pi/2)}$.
If $k=(N-n)/2$, then $\alpha-2\pi k/N=\pi$ and it follows that
$g_{00}/g_{01}=(-1)^ze^{i(-\alpha+\beta+\pi/2)}$.
The eigenstates are thus equal superpositions of the coin components
$\ket{0}$ and $\ket{1}$, with the phase difference determined only by
the coin parameters $\alpha$ and $\beta$ (and the eigenvalue label $z$).
This property is reflected in Fig. \ref{fig:ominaisvektorit}b)
where the Bloch angle $\theta$ at $k=0,N/2$ is equal to $\pi/2$ for
all values of $R$.
%TODO determine the lowest and highest eigenvalues

The robustness of the special eigenstates might be used to
store quantum states and to protect them against arbitrary variations
of the parameter $R$.
If the eigenvalues $z=1,2$ at wavenumber $k$ are unique and the rest of
the spectrum is degenerate, the eigenstates $\ket{\psi_{\lambda(k,1)}}$
and $\ket{\psi_{\lambda(k,2)}}$ remain ``frozen'' even if the parameter
$R$ changes at every time step, as long as $R$ is spatially homogeneous
and the parameters $\alpha$ and $\beta$ remain fixed.
Therefore, if the initial state is
\begin{equation}
\ket{\psi(0)} = x_0 \ket{\varphi}
+ x_1 \ket{\psi_{\lambda(k,1)}} + x_2 \ket{\psi_{\lambda(k,2)}}
\end{equation}
where $\ket{\varphi}$ is orthogonal to $\ket{\psi_{\lambda(k,1)}}$
and $\ket{\psi_{\lambda(k,2)}}$, the overlaps between
the state at time $t$ and the eigenstates remain fixed at all times,
up to a complex phase:
\begin{align}
\big\langle\psi_{\lambda(k,1)}\ket{\psi(t)} \; = \; & e^{iy_1}x_1 \\
\big\langle\psi_{\lambda(k,2)}\ket{\psi(t)} \; = \; & e^{iy_2}x_2,
\end{align}
and the probabilities to measure the states $\ket{\psi_{\lambda(k,1)}}$
and $\ket{\psi_{\lambda(k,2)}}$ are constant.
The quantum walk can thus act as a memory which is protected from
temporal variations of $R$.

The robustness of the protected states is reminiscent of symmetry-protected
topological phases, where the edges between different phases may host chiral
states inside the band gap.
Such states are also robust to arbitrary variations of the system parameters,
as long as the symmetry class of the Hamiltonian does not change.
Such states have also been found in the so-called split-step quantum walk
\cite{krbd2010,kitagawa2012} and also in disordered quantum walks \cite{ok2011}.
In fact, the topological structure of quantum walks has been found to be
richer than that of non-interacting many-body systems, and the topological
features in the quantum walk are not always captured by standard tools
used to characterize topological phases in non-interacting systems
\cite{asboth2012}.
In the simple case that we have considered here, the properties of the
eigenstates were found to be very different in the presence of symmetry.
The topological features in quantum walks are still not completely
understood, and it interesting that even the simplest quantum
walk can host states which are robust against certain variations
of the system parameters.

\section{Conclusions} \label{sec:johtopaatokset}

The spectrum of the time evolution operator of discrete-time
quantum walks was analysed for a general
2-dimensional unitary coin.
The eigenvalues are labelled by two integers $k$ and $z$,
where $k$ is the wavenumber in Fourier space and $z$ labels
the upper and lower quasi-energy bands.
If the coin is parameterized suitably, each parameter can be
associated with distinct properties of the spectrum.
The bias parameter $R$ controls the dispersion of the bands,
the angle $\alpha$ changes the symmetry between left-moving and
right-moving components, and the angle $\beta$ sets just the
base level of the quasi-energies.

For the eigenstates, the geometric picture using the Bloch ball
was found to be very useful.
The reduced eigenvectors in coin space are distributed along well-defined
trajectories, which become deformed as the coin parameters are varied.
For small values of the bias parameter $R$, the eigenvectors
remain close to the equator.
Increasing $R$ brings them closer to the poles, with all eigenvectors
localized at either $c=0$ or $c=1$ when $R=1$.
If the parameter $\alpha$ is an integer multiple $\pi/N$, the spectrum
becomes degenerate, which has a significant impact on the eigenstates.
In the non-degenerate case, the eigenvectors are product states of
the Fourier modes and coin states, but in the presence of degeneracy they
are entangled superpositions of conjugate wavenumbers $k$ and $k'$.
In the latter case the reduced coin states are mixed states, and
they are not constrained to the surface of the Bloch sphere.
The representation of the eigenvectors is also not unique in this case,
and we have identified the parameters $s_1$ and $\omega_1$ as
a certain kind gauge freedom, which can be chosen freely while the
resulting states are still orthonormal eigenstates of $U$.
However, the Bloch representation of the eigenvectors is not equal
for different $s_1$ and $\omega_1$.

Finally, depending on the evenness of the number of lattices sites $N$
and the difference $(N-n)$, where $\alpha=n\pi/N$, it has been shown
that there may exist unique eigenvalues in the degenerate spectrum.
These points are located at the top or bottom of the quasi-energy
bands, and the eigenstates corresponding to the unique eigenvalues
are robust against arbitrary variations in the parameter $R$.
If the initial state of the quantum walk has a non-zero overlap
with such states, the overlap will remain constant in time, even
if the parameter $R$ changes between time steps.
Since these states remain constant in time, they might be useful for storing
quantum states, and such a memory would be protected from arbitrary
temporal variations in the parameter $R$.

\vspace{4ex}
\noindent
\emph{Acknowledgments}

The author acknowledges support by the Alexander von Humboldt foundation.

% reunat ja wavelet-kanta
% jatkuvuusraja: N->aareton kun ympyran keha pysyy vakiona ja paikkojen valiset etaisyydet->0
% (ks. andrew childs 2010)
% satunnaismatriisien kaytto epahomog. kolikoilla: lokalisoituminen / ymparisto

% \bibliographystyle{abbrv}
\bibliographystyle{qwspectrum}
\bibliography{./references} % run bibtex articletemplate.aux

\end{document}

%% file: ominaisarvot_r.tex
% GNUPLOT: LaTeX picture with Postscript
\begingroup
  \makeatletter
  \providecommand\color[2][]{%
    \GenericError{(gnuplot) \space\space\space\@spaces}{%
      Package color not loaded in conjunction with
      terminal option `colourtext'%
    }{See the gnuplot documentation for explanation.%
    }{Either use 'blacktext' in gnuplot or load the package
      color.sty in LaTeX.}%
    \renewcommand\color[2][]{}%
  }%
  \providecommand\includegraphics[2][]{%
    \GenericError{(gnuplot) \space\space\space\@spaces}{%
      Package graphicx or graphics not loaded%
    }{See the gnuplot documentation for explanation.%
    }{The gnuplot epslatex terminal needs graphicx.sty or graphics.sty.}%
    \renewcommand\includegraphics[2][]{}%
  }%
  \providecommand\rotatebox[2]{#2}%
  \@ifundefined{ifGPcolor}{%
    \newif\ifGPcolor
    \GPcolortrue
  }{}%
  \@ifundefined{ifGPblacktext}{%
    \newif\ifGPblacktext
    \GPblacktexttrue
  }{}%
  % define a \g@addto@macro without @ in the name:
  \let\gplgaddtomacro\g@addto@macro
  % define empty templates for all commands taking text:
  \gdef\gplbacktext{}%
  \gdef\gplfronttext{}%
  \makeatother
  \ifGPblacktext
    % no textcolor at all
    \def\colorrgb#1{}%
    \def\colorgray#1{}%
  \else
    % gray or color?
    \ifGPcolor
      \def\colorrgb#1{\color[rgb]{#1}}%
      \def\colorgray#1{\color[gray]{#1}}%
      \expandafter\def\csname LTw\endcsname{\color{white}}%
      \expandafter\def\csname LTb\endcsname{\color{black}}%
      \expandafter\def\csname LTa\endcsname{\color{black}}%
      \expandafter\def\csname LT0\endcsname{\color[rgb]{1,0,0}}%
      \expandafter\def\csname LT1\endcsname{\color[rgb]{0,1,0}}%
      \expandafter\def\csname LT2\endcsname{\color[rgb]{0,0,1}}%
      \expandafter\def\csname LT3\endcsname{\color[rgb]{1,0,1}}%
      \expandafter\def\csname LT4\endcsname{\color[rgb]{0,1,1}}%
      \expandafter\def\csname LT5\endcsname{\color[rgb]{1,1,0}}%
      \expandafter\def\csname LT6\endcsname{\color[rgb]{0,0,0}}%
      \expandafter\def\csname LT7\endcsname{\color[rgb]{1,0.3,0}}%
      \expandafter\def\csname LT8\endcsname{\color[rgb]{0.5,0.5,0.5}}%
    \else
      % gray
      \def\colorrgb#1{\color{black}}%
      \def\colorgray#1{\color[gray]{#1}}%
      \expandafter\def\csname LTw\endcsname{\color{white}}%
      \expandafter\def\csname LTb\endcsname{\color{black}}%
      \expandafter\def\csname LTa\endcsname{\color{black}}%
      \expandafter\def\csname LT0\endcsname{\color{black}}%
      \expandafter\def\csname LT1\endcsname{\color{black}}%
      \expandafter\def\csname LT2\endcsname{\color{black}}%
      \expandafter\def\csname LT3\endcsname{\color{black}}%
      \expandafter\def\csname LT4\endcsname{\color{black}}%
      \expandafter\def\csname LT5\endcsname{\color{black}}%
      \expandafter\def\csname LT6\endcsname{\color{black}}%
      \expandafter\def\csname LT7\endcsname{\color{black}}%
      \expandafter\def\csname LT8\endcsname{\color{black}}%
    \fi
  \fi
  \setlength{\unitlength}{0.0500bp}%
  \begin{picture}(3744.00,2590.00)%
    \gplgaddtomacro\gplbacktext{%
      \csname LTb\endcsname%
      \put(528,550){\makebox(0,0)[r]{\strut{}\scriptsize -1}}%
      \put(528,635){\makebox(0,0)[r]{\strut{}}}%
      \put(528,721){\makebox(0,0)[r]{\strut{}}}%
      \put(528,806){\makebox(0,0)[r]{\strut{}}}%
      \put(528,892){\makebox(0,0)[r]{\strut{}}}%
      \put(528,977){\makebox(0,0)[r]{\strut{}\scriptsize -0.5}}%
      \put(528,1063){\makebox(0,0)[r]{\strut{}}}%
      \put(528,1148){\makebox(0,0)[r]{\strut{}}}%
      \put(528,1234){\makebox(0,0)[r]{\strut{}}}%
      \put(528,1319){\makebox(0,0)[r]{\strut{}}}%
      \put(528,1405){\makebox(0,0)[r]{\strut{}\scriptsize 0}}%
      \put(528,1490){\makebox(0,0)[r]{\strut{}}}%
      \put(528,1575){\makebox(0,0)[r]{\strut{}}}%
      \put(528,1661){\makebox(0,0)[r]{\strut{}}}%
      \put(528,1746){\makebox(0,0)[r]{\strut{}}}%
      \put(528,1832){\makebox(0,0)[r]{\strut{}\scriptsize 0.5}}%
      \put(528,1917){\makebox(0,0)[r]{\strut{}}}%
      \put(528,2003){\makebox(0,0)[r]{\strut{}}}%
      \put(528,2088){\makebox(0,0)[r]{\strut{}}}%
      \put(528,2174){\makebox(0,0)[r]{\strut{}}}%
      \put(528,2259){\makebox(0,0)[r]{\strut{}\scriptsize 1}}%
      \put(660,330){\makebox(0,0){\strut{}\scriptsize 0}}%
      \put(955,330){\makebox(0,0){\strut{}}}%
      \put(1250,330){\makebox(0,0){\strut{}\scriptsize 0.2}}%
      \put(1545,330){\makebox(0,0){\strut{}}}%
      \put(1840,330){\makebox(0,0){\strut{}\scriptsize 0.4}}%
      \put(2136,330){\makebox(0,0){\strut{}}}%
      \put(2431,330){\makebox(0,0){\strut{}\scriptsize 0.6}}%
      \put(2726,330){\makebox(0,0){\strut{}}}%
      \put(3021,330){\makebox(0,0){\strut{}\scriptsize 0.8}}%
      \put(3316,330){\makebox(0,0){\strut{}}}%
      \put(3611,330){\makebox(0,0){\strut{}\scriptsize 1}}%
      \put(88,1404){\rotatebox{-270}{\makebox(0,0){\strut{}\small $\lambda/\pi$}}}%
      \put(2135,110){\makebox(0,0){\strut{}\small $k/N$}}%
      \put(2135,2435){\makebox(0,0){\strut{}a)}}%
    }%
    \gplgaddtomacro\gplfronttext{%
      \csname LTb\endcsname%
      \put(2938,2145){\makebox(0,0)[l]{\strut{}\tiny$R$=0}}%
      \csname LTb\endcsname%
      \put(2938,2002){\makebox(0,0)[l]{\strut{}\tiny$R$=0.25}}%
      \csname LTb\endcsname%
      \put(2938,1859){\makebox(0,0)[l]{\strut{}\tiny$R$=0.5}}%
      \csname LTb\endcsname%
      \put(2938,1716){\makebox(0,0)[l]{\strut{}\tiny$R$=0.75}}%
      \csname LTb\endcsname%
      \put(2938,1573){\makebox(0,0)[l]{\strut{}\tiny$R$=1}}%
    }%
    \gplbacktext
    \put(0,0){\includegraphics{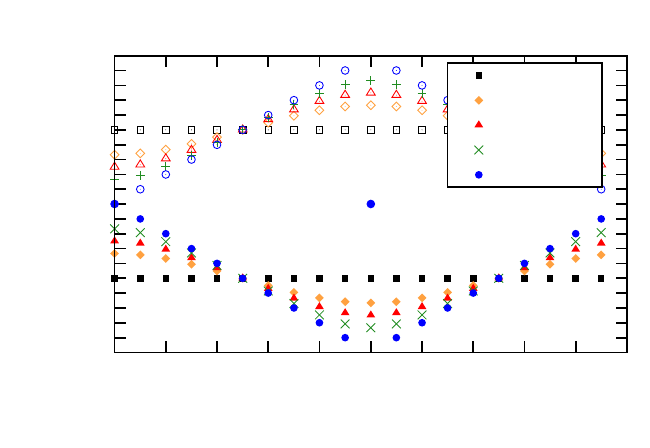}}%
    \gplfronttext
  \end{picture}%
\endgroup

%% file: ominaisarvot_alphabeta.tex
% GNUPLOT: LaTeX picture with Postscript
\begingroup
  \makeatletter
  \providecommand\color[2][]{%
    \GenericError{(gnuplot) \space\space\space\@spaces}{%
      Package color not loaded in conjunction with
      terminal option `colourtext'%
    }{See the gnuplot documentation for explanation.%
    }{Either use 'blacktext' in gnuplot or load the package
      color.sty in LaTeX.}%
    \renewcommand\color[2][]{}%
  }%
  \providecommand\includegraphics[2][]{%
    \GenericError{(gnuplot) \space\space\space\@spaces}{%
      Package graphicx or graphics not loaded%
    }{See the gnuplot documentation for explanation.%
    }{The gnuplot epslatex terminal needs graphicx.sty or graphics.sty.}%
    \renewcommand\includegraphics[2][]{}%
  }%
  \providecommand\rotatebox[2]{#2}%
  \@ifundefined{ifGPcolor}{%
    \newif\ifGPcolor
    \GPcolortrue
  }{}%
  \@ifundefined{ifGPblacktext}{%
    \newif\ifGPblacktext
    \GPblacktexttrue
  }{}%
  % define a \g@addto@macro without @ in the name:
  \let\gplgaddtomacro\g@addto@macro
  % define empty templates for all commands taking text:
  \gdef\gplbacktext{}%
  \gdef\gplfronttext{}%
  \makeatother
  \ifGPblacktext
    % no textcolor at all
    \def\colorrgb#1{}%
    \def\colorgray#1{}%
  \else
    % gray or color?
    \ifGPcolor
      \def\colorrgb#1{\color[rgb]{#1}}%
      \def\colorgray#1{\color[gray]{#1}}%
      \expandafter\def\csname LTw\endcsname{\color{white}}%
      \expandafter\def\csname LTb\endcsname{\color{black}}%
      \expandafter\def\csname LTa\endcsname{\color{black}}%
      \expandafter\def\csname LT0\endcsname{\color[rgb]{1,0,0}}%
      \expandafter\def\csname LT1\endcsname{\color[rgb]{0,1,0}}%
      \expandafter\def\csname LT2\endcsname{\color[rgb]{0,0,1}}%
      \expandafter\def\csname LT3\endcsname{\color[rgb]{1,0,1}}%
      \expandafter\def\csname LT4\endcsname{\color[rgb]{0,1,1}}%
      \expandafter\def\csname LT5\endcsname{\color[rgb]{1,1,0}}%
      \expandafter\def\csname LT6\endcsname{\color[rgb]{0,0,0}}%
      \expandafter\def\csname LT7\endcsname{\color[rgb]{1,0.3,0}}%
      \expandafter\def\csname LT8\endcsname{\color[rgb]{0.5,0.5,0.5}}%
    \else
      % gray
      \def\colorrgb#1{\color{black}}%
      \def\colorgray#1{\color[gray]{#1}}%
      \expandafter\def\csname LTw\endcsname{\color{white}}%
      \expandafter\def\csname LTb\endcsname{\color{black}}%
      \expandafter\def\csname LTa\endcsname{\color{black}}%
      \expandafter\def\csname LT0\endcsname{\color{black}}%
      \expandafter\def\csname LT1\endcsname{\color{black}}%
      \expandafter\def\csname LT2\endcsname{\color{black}}%
      \expandafter\def\csname LT3\endcsname{\color{black}}%
      \expandafter\def\csname LT4\endcsname{\color{black}}%
      \expandafter\def\csname LT5\endcsname{\color{black}}%
      \expandafter\def\csname LT6\endcsname{\color{black}}%
      \expandafter\def\csname LT7\endcsname{\color{black}}%
      \expandafter\def\csname LT8\endcsname{\color{black}}%
    \fi
  \fi
  \setlength{\unitlength}{0.0500bp}%
  \begin{picture}(2880.00,2590.00)%
    \gplgaddtomacro\gplbacktext{%
      \csname LTb\endcsname%
      \put(528,550){\makebox(0,0)[r]{\strut{}\scriptsize -1}}%
      \put(528,635){\makebox(0,0)[r]{\strut{}}}%
      \put(528,721){\makebox(0,0)[r]{\strut{}}}%
      \put(528,806){\makebox(0,0)[r]{\strut{}}}%
      \put(528,892){\makebox(0,0)[r]{\strut{}}}%
      \put(528,977){\makebox(0,0)[r]{\strut{}\scriptsize -0.5}}%
      \put(528,1063){\makebox(0,0)[r]{\strut{}}}%
      \put(528,1148){\makebox(0,0)[r]{\strut{}}}%
      \put(528,1234){\makebox(0,0)[r]{\strut{}}}%
      \put(528,1319){\makebox(0,0)[r]{\strut{}}}%
      \put(528,1405){\makebox(0,0)[r]{\strut{}\scriptsize 0}}%
      \put(528,1490){\makebox(0,0)[r]{\strut{}}}%
      \put(528,1575){\makebox(0,0)[r]{\strut{}}}%
      \put(528,1661){\makebox(0,0)[r]{\strut{}}}%
      \put(528,1746){\makebox(0,0)[r]{\strut{}}}%
      \put(528,1832){\makebox(0,0)[r]{\strut{}\scriptsize 0.5}}%
      \put(528,1917){\makebox(0,0)[r]{\strut{}}}%
      \put(528,2003){\makebox(0,0)[r]{\strut{}}}%
      \put(528,2088){\makebox(0,0)[r]{\strut{}}}%
      \put(528,2174){\makebox(0,0)[r]{\strut{}}}%
      \put(528,2259){\makebox(0,0)[r]{\strut{}\scriptsize 1}}%
      \put(660,330){\makebox(0,0){\strut{}\scriptsize 0}}%
      \put(869,330){\makebox(0,0){\strut{}}}%
      \put(1077,330){\makebox(0,0){\strut{}\scriptsize 0.2}}%
      \put(1286,330){\makebox(0,0){\strut{}}}%
      \put(1495,330){\makebox(0,0){\strut{}\scriptsize 0.4}}%
      \put(1704,330){\makebox(0,0){\strut{}}}%
      \put(1912,330){\makebox(0,0){\strut{}\scriptsize 0.6}}%
      \put(2121,330){\makebox(0,0){\strut{}}}%
      \put(2330,330){\makebox(0,0){\strut{}\scriptsize 0.8}}%
      \put(2538,330){\makebox(0,0){\strut{}}}%
      \put(2747,330){\makebox(0,0){\strut{}\scriptsize 1}}%
      \put(88,1404){\rotatebox{-270}{\makebox(0,0){\strut{}\small $\lambda/\pi$}}}%
      \put(1703,110){\makebox(0,0){\strut{}\small $k/N$}}%
      \put(1703,2435){\makebox(0,0){\strut{}b)}}%
    }%
    \gplgaddtomacro\gplfronttext{%
      \csname LTb\endcsname%
      \put(1917,1718){\makebox(0,0)[l]{\strut{}\tiny$\alpha=0$}}%
      \csname LTb\endcsname%
      \put(1917,1575){\makebox(0,0)[l]{\strut{}\tiny$\alpha=-\pi/N$}}%
      \csname LTb\endcsname%
      \put(1917,1432){\makebox(0,0)[l]{\strut{}\tiny$\alpha=\pi/N$}}%
      \csname LTb\endcsname%
      \put(1917,1289){\makebox(0,0)[l]{\strut{}\tiny$\beta=0$}}%
      \csname LTb\endcsname%
      \put(1917,1146){\makebox(0,0)[l]{\strut{}\tiny$\beta=\pi/N$}}%
    }%
    \gplbacktext
    \put(0,0){\includegraphics{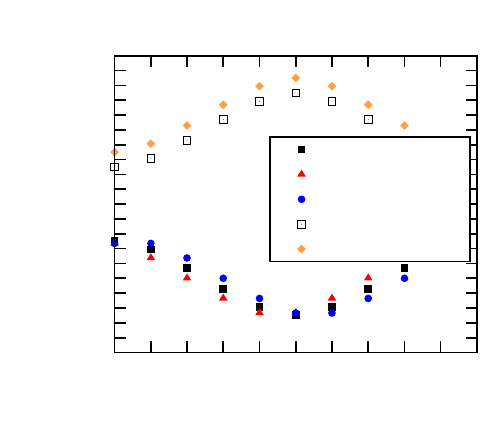}}%
    \gplfronttext
  \end{picture}%
\endgroup

%% file: ominvekt_r.tex
% GNUPLOT: LaTeX picture with Postscript
\begingroup
  \makeatletter
  \providecommand\color[2][]{%
    \GenericError{(gnuplot) \space\space\space\@spaces}{%
      Package color not loaded in conjunction with
      terminal option `colourtext'%
    }{See the gnuplot documentation for explanation.%
    }{Either use 'blacktext' in gnuplot or load the package
      color.sty in LaTeX.}%
    \renewcommand\color[2][]{}%
  }%
  \providecommand\includegraphics[2][]{%
    \GenericError{(gnuplot) \space\space\space\@spaces}{%
      Package graphicx or graphics not loaded%
    }{See the gnuplot documentation for explanation.%
    }{The gnuplot epslatex terminal needs graphicx.sty or graphics.sty.}%
    \renewcommand\includegraphics[2][]{}%
  }%
  \providecommand\rotatebox[2]{#2}%
  \@ifundefined{ifGPcolor}{%
    \newif\ifGPcolor
    \GPcolortrue
  }{}%
  \@ifundefined{ifGPblacktext}{%
    \newif\ifGPblacktext
    \GPblacktexttrue
  }{}%
  % define a \g@addto@macro without @ in the name:
  \let\gplgaddtomacro\g@addto@macro
  % define empty templates for all commands taking text:
  \gdef\gplbacktext{}%
  \gdef\gplfronttext{}%
  \makeatother
  \ifGPblacktext
    % no textcolor at all
    \def\colorrgb#1{}%
    \def\colorgray#1{}%
  \else
    % gray or color?
    \ifGPcolor
      \def\colorrgb#1{\color[rgb]{#1}}%
      \def\colorgray#1{\color[gray]{#1}}%
      \expandafter\def\csname LTw\endcsname{\color{white}}%
      \expandafter\def\csname LTb\endcsname{\color{black}}%
      \expandafter\def\csname LTa\endcsname{\color{black}}%
      \expandafter\def\csname LT0\endcsname{\color[rgb]{1,0,0}}%
      \expandafter\def\csname LT1\endcsname{\color[rgb]{0,1,0}}%
      \expandafter\def\csname LT2\endcsname{\color[rgb]{0,0,1}}%
      \expandafter\def\csname LT3\endcsname{\color[rgb]{1,0,1}}%
      \expandafter\def\csname LT4\endcsname{\color[rgb]{0,1,1}}%
      \expandafter\def\csname LT5\endcsname{\color[rgb]{1,1,0}}%
      \expandafter\def\csname LT6\endcsname{\color[rgb]{0,0,0}}%
      \expandafter\def\csname LT7\endcsname{\color[rgb]{1,0.3,0}}%
      \expandafter\def\csname LT8\endcsname{\color[rgb]{0.5,0.5,0.5}}%
    \else
      % gray
      \def\colorrgb#1{\color{black}}%
      \def\colorgray#1{\color[gray]{#1}}%
      \expandafter\def\csname LTw\endcsname{\color{white}}%
      \expandafter\def\csname LTb\endcsname{\color{black}}%
      \expandafter\def\csname LTa\endcsname{\color{black}}%
      \expandafter\def\csname LT0\endcsname{\color{black}}%
      \expandafter\def\csname LT1\endcsname{\color{black}}%
      \expandafter\def\csname LT2\endcsname{\color{black}}%
      \expandafter\def\csname LT3\endcsname{\color{black}}%
      \expandafter\def\csname LT4\endcsname{\color{black}}%
      \expandafter\def\csname LT5\endcsname{\color{black}}%
      \expandafter\def\csname LT6\endcsname{\color{black}}%
      \expandafter\def\csname LT7\endcsname{\color{black}}%
      \expandafter\def\csname LT8\endcsname{\color{black}}%
    \fi
  \fi
  \setlength{\unitlength}{0.0500bp}%
  \begin{picture}(3744.00,2880.00)%
    \gplgaddtomacro\gplbacktext{%
      \csname LTb\endcsname%
      \put(528,550){\makebox(0,0)[r]{\strut{}\scriptsize 0}}%
      \put(528,728){\makebox(0,0)[r]{\strut{}}}%
      \put(528,906){\makebox(0,0)[r]{\strut{}\scriptsize 0.2}}%
      \put(528,1084){\makebox(0,0)[r]{\strut{}}}%
      \put(528,1262){\makebox(0,0)[r]{\strut{}\scriptsize 0.4}}%
      \put(528,1440){\makebox(0,0)[r]{\strut{}}}%
      \put(528,1617){\makebox(0,0)[r]{\strut{}\scriptsize 0.6}}%
      \put(528,1795){\makebox(0,0)[r]{\strut{}}}%
      \put(528,1973){\makebox(0,0)[r]{\strut{}\scriptsize 0.8}}%
      \put(528,2151){\makebox(0,0)[r]{\strut{}}}%
      \put(528,2329){\makebox(0,0)[r]{\strut{}\scriptsize 1}}%
      \put(660,330){\makebox(0,0){\strut{}\scriptsize 0}}%
      \put(942,330){\makebox(0,0){\strut{}}}%
      \put(1224,330){\makebox(0,0){\strut{}\scriptsize 0.2}}%
      \put(1506,330){\makebox(0,0){\strut{}}}%
      \put(1788,330){\makebox(0,0){\strut{}\scriptsize 0.4}}%
      \put(2070,330){\makebox(0,0){\strut{}}}%
      \put(2351,330){\makebox(0,0){\strut{}\scriptsize 0.6}}%
      \put(2633,330){\makebox(0,0){\strut{}}}%
      \put(2915,330){\makebox(0,0){\strut{}\scriptsize 0.8}}%
      \put(3197,330){\makebox(0,0){\strut{}}}%
      \put(3479,330){\makebox(0,0){\strut{}\scriptsize 1}}%
      \put(88,1439){\rotatebox{-270}{\makebox(0,0){\strut{}\small $r$}}}%
      \put(2069,2527){\makebox(0,0){\strut{}\small a)}}%
    }%
    \gplgaddtomacro\gplfronttext{%
      \csname LTb\endcsname%
      \put(2852,1315){\makebox(0,0)[l]{\strut{}\tiny$R$=0}}%
      \csname LTb\endcsname%
      \put(2852,1172){\makebox(0,0)[l]{\strut{}\tiny$R$=0.25}}%
      \csname LTb\endcsname%
      \put(2852,1029){\makebox(0,0)[l]{\strut{}\tiny$R$=0.5}}%
      \csname LTb\endcsname%
      \put(2852,886){\makebox(0,0)[l]{\strut{}\tiny$R$=0.75}}%
      \csname LTb\endcsname%
      \put(2852,743){\makebox(0,0)[l]{\strut{}\tiny$R$=.95}}%
      \csname LTb\endcsname%
      \put(2852,600){\makebox(0,0)[l]{\strut{}\tiny$R$=1}}%
    }%
    \gplbacktext
    \put(0,0){\includegraphics{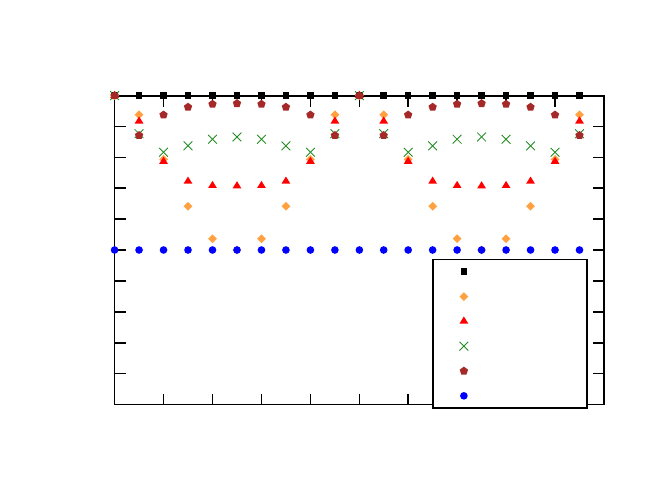}}%
    \gplfronttext
  \end{picture}%
\endgroup

%% file: blochpallo.tex
\begin{picture}(.65,.6)
% \put(0,0){\line(0,1){.6}}
% \put(0,0){\line(1,0){.65}}
% \put(.65,0){\line(0,1){.6}}
% \put(0,.6){\line(1,0){.65}}
\put(.05,.07){\includegraphics[width=.5\unitlength]{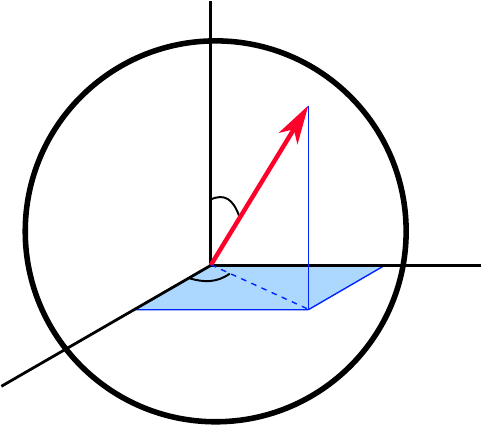}}
\put(.01,.08){\tiny $x$}
\put(.57,.23){\tiny $y$}
\put(.24,.52){\tiny $z$}
\put(.32,.285){\tiny $r$}
\put(.285,.355){\tiny $\theta$}
\put(.25,.17){\tiny $\phi$}
\put(.29,.49){\small $|0\rangle$}
\put(.25,.0){\small $|1\rangle$}
\end{picture}

%% file: ominvekt_theta.tex
% GNUPLOT: LaTeX picture with Postscript
\begingroup
  \makeatletter
  \providecommand\color[2][]{%
    \GenericError{(gnuplot) \space\space\space\@spaces}{%
      Package color not loaded in conjunction with
      terminal option `colourtext'%
    }{See the gnuplot documentation for explanation.%
    }{Either use 'blacktext' in gnuplot or load the package
      color.sty in LaTeX.}%
    \renewcommand\color[2][]{}%
  }%
  \providecommand\includegraphics[2][]{%
    \GenericError{(gnuplot) \space\space\space\@spaces}{%
      Package graphicx or graphics not loaded%
    }{See the gnuplot documentation for explanation.%
    }{The gnuplot epslatex terminal needs graphicx.sty or graphics.sty.}%
    \renewcommand\includegraphics[2][]{}%
  }%
  \providecommand\rotatebox[2]{#2}%
  \@ifundefined{ifGPcolor}{%
    \newif\ifGPcolor
    \GPcolortrue
  }{}%
  \@ifundefined{ifGPblacktext}{%
    \newif\ifGPblacktext
    \GPblacktexttrue
  }{}%
  % define a \g@addto@macro without @ in the name:
  \let\gplgaddtomacro\g@addto@macro
  % define empty templates for all commands taking text:
  \gdef\gplbacktext{}%
  \gdef\gplfronttext{}%
  \makeatother
  \ifGPblacktext
    % no textcolor at all
    \def\colorrgb#1{}%
    \def\colorgray#1{}%
  \else
    % gray or color?
    \ifGPcolor
      \def\colorrgb#1{\color[rgb]{#1}}%
      \def\colorgray#1{\color[gray]{#1}}%
      \expandafter\def\csname LTw\endcsname{\color{white}}%
      \expandafter\def\csname LTb\endcsname{\color{black}}%
      \expandafter\def\csname LTa\endcsname{\color{black}}%
      \expandafter\def\csname LT0\endcsname{\color[rgb]{1,0,0}}%
      \expandafter\def\csname LT1\endcsname{\color[rgb]{0,1,0}}%
      \expandafter\def\csname LT2\endcsname{\color[rgb]{0,0,1}}%
      \expandafter\def\csname LT3\endcsname{\color[rgb]{1,0,1}}%
      \expandafter\def\csname LT4\endcsname{\color[rgb]{0,1,1}}%
      \expandafter\def\csname LT5\endcsname{\color[rgb]{1,1,0}}%
      \expandafter\def\csname LT6\endcsname{\color[rgb]{0,0,0}}%
      \expandafter\def\csname LT7\endcsname{\color[rgb]{1,0.3,0}}%
      \expandafter\def\csname LT8\endcsname{\color[rgb]{0.5,0.5,0.5}}%
    \else
      % gray
      \def\colorrgb#1{\color{black}}%
      \def\colorgray#1{\color[gray]{#1}}%
      \expandafter\def\csname LTw\endcsname{\color{white}}%
      \expandafter\def\csname LTb\endcsname{\color{black}}%
      \expandafter\def\csname LTa\endcsname{\color{black}}%
      \expandafter\def\csname LT0\endcsname{\color{black}}%
      \expandafter\def\csname LT1\endcsname{\color{black}}%
      \expandafter\def\csname LT2\endcsname{\color{black}}%
      \expandafter\def\csname LT3\endcsname{\color{black}}%
      \expandafter\def\csname LT4\endcsname{\color{black}}%
      \expandafter\def\csname LT5\endcsname{\color{black}}%
      \expandafter\def\csname LT6\endcsname{\color{black}}%
      \expandafter\def\csname LT7\endcsname{\color{black}}%
      \expandafter\def\csname LT8\endcsname{\color{black}}%
    \fi
  \fi
  \setlength{\unitlength}{0.0500bp}%
  \begin{picture}(3744.00,2880.00)%
    \gplgaddtomacro\gplbacktext{%
      \csname LTb\endcsname%
      \put(528,550){\makebox(0,0)[r]{\strut{}\scriptsize 0}}%
      \put(528,728){\makebox(0,0)[r]{\strut{}}}%
      \put(528,906){\makebox(0,0)[r]{\strut{}\scriptsize 0.2}}%
      \put(528,1084){\makebox(0,0)[r]{\strut{}}}%
      \put(528,1262){\makebox(0,0)[r]{\strut{}\scriptsize 0.4}}%
      \put(528,1440){\makebox(0,0)[r]{\strut{}}}%
      \put(528,1617){\makebox(0,0)[r]{\strut{}\scriptsize 0.6}}%
      \put(528,1795){\makebox(0,0)[r]{\strut{}}}%
      \put(528,1973){\makebox(0,0)[r]{\strut{}\scriptsize 0.8}}%
      \put(528,2151){\makebox(0,0)[r]{\strut{}}}%
      \put(528,2329){\makebox(0,0)[r]{\strut{}\scriptsize 1}}%
      \put(660,330){\makebox(0,0){\strut{}\scriptsize 0}}%
      \put(942,330){\makebox(0,0){\strut{}}}%
      \put(1224,330){\makebox(0,0){\strut{}\scriptsize 0.2}}%
      \put(1506,330){\makebox(0,0){\strut{}}}%
      \put(1788,330){\makebox(0,0){\strut{}\scriptsize 0.4}}%
      \put(2070,330){\makebox(0,0){\strut{}}}%
      \put(2351,330){\makebox(0,0){\strut{}\scriptsize 0.6}}%
      \put(2633,330){\makebox(0,0){\strut{}}}%
      \put(2915,330){\makebox(0,0){\strut{}\scriptsize 0.8}}%
      \put(3197,330){\makebox(0,0){\strut{}}}%
      \put(3479,330){\makebox(0,0){\strut{}\scriptsize 1}}%
      \put(88,1439){\rotatebox{-270}{\makebox(0,0){\strut{}\small $\theta/\pi$}}}%
      \put(2069,110){\makebox(0,0){\strut{}\small $k/N$}}%
      \put(2069,2527){\makebox(0,0){\strut{}\small b)}}%
    }%
    \gplgaddtomacro\gplfronttext{%
    }%
    \gplbacktext
    \put(0,0){\includegraphics{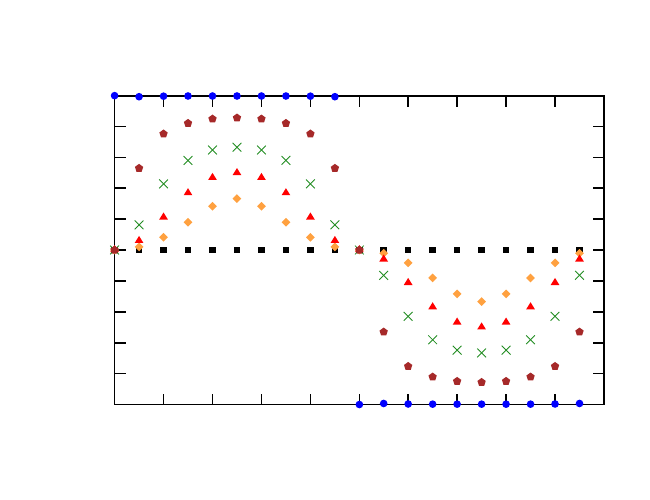}}%
    \gplfronttext
  \end{picture}%
\endgroup

%% file: ominvekt_phi.tex
% GNUPLOT: LaTeX picture with Postscript
\begingroup
  \makeatletter
  \providecommand\color[2][]{%
    \GenericError{(gnuplot) \space\space\space\@spaces}{%
      Package color not loaded in conjunction with
      terminal option `colourtext'%
    }{See the gnuplot documentation for explanation.%
    }{Either use 'blacktext' in gnuplot or load the package
      color.sty in LaTeX.}%
    \renewcommand\color[2][]{}%
  }%
  \providecommand\includegraphics[2][]{%
    \GenericError{(gnuplot) \space\space\space\@spaces}{%
      Package graphicx or graphics not loaded%
    }{See the gnuplot documentation for explanation.%
    }{The gnuplot epslatex terminal needs graphicx.sty or graphics.sty.}%
    \renewcommand\includegraphics[2][]{}%
  }%
  \providecommand\rotatebox[2]{#2}%
  \@ifundefined{ifGPcolor}{%
    \newif\ifGPcolor
    \GPcolortrue
  }{}%
  \@ifundefined{ifGPblacktext}{%
    \newif\ifGPblacktext
    \GPblacktexttrue
  }{}%
  % define a \g@addto@macro without @ in the name:
  \let\gplgaddtomacro\g@addto@macro
  % define empty templates for all commands taking text:
  \gdef\gplbacktext{}%
  \gdef\gplfronttext{}%
  \makeatother
  \ifGPblacktext
    % no textcolor at all
    \def\colorrgb#1{}%
    \def\colorgray#1{}%
  \else
    % gray or color?
    \ifGPcolor
      \def\colorrgb#1{\color[rgb]{#1}}%
      \def\colorgray#1{\color[gray]{#1}}%
      \expandafter\def\csname LTw\endcsname{\color{white}}%
      \expandafter\def\csname LTb\endcsname{\color{black}}%
      \expandafter\def\csname LTa\endcsname{\color{black}}%
      \expandafter\def\csname LT0\endcsname{\color[rgb]{1,0,0}}%
      \expandafter\def\csname LT1\endcsname{\color[rgb]{0,1,0}}%
      \expandafter\def\csname LT2\endcsname{\color[rgb]{0,0,1}}%
      \expandafter\def\csname LT3\endcsname{\color[rgb]{1,0,1}}%
      \expandafter\def\csname LT4\endcsname{\color[rgb]{0,1,1}}%
      \expandafter\def\csname LT5\endcsname{\color[rgb]{1,1,0}}%
      \expandafter\def\csname LT6\endcsname{\color[rgb]{0,0,0}}%
      \expandafter\def\csname LT7\endcsname{\color[rgb]{1,0.3,0}}%
      \expandafter\def\csname LT8\endcsname{\color[rgb]{0.5,0.5,0.5}}%
    \else
      % gray
      \def\colorrgb#1{\color{black}}%
      \def\colorgray#1{\color[gray]{#1}}%
      \expandafter\def\csname LTw\endcsname{\color{white}}%
      \expandafter\def\csname LTb\endcsname{\color{black}}%
      \expandafter\def\csname LTa\endcsname{\color{black}}%
      \expandafter\def\csname LT0\endcsname{\color{black}}%
      \expandafter\def\csname LT1\endcsname{\color{black}}%
      \expandafter\def\csname LT2\endcsname{\color{black}}%
      \expandafter\def\csname LT3\endcsname{\color{black}}%
      \expandafter\def\csname LT4\endcsname{\color{black}}%
      \expandafter\def\csname LT5\endcsname{\color{black}}%
      \expandafter\def\csname LT6\endcsname{\color{black}}%
      \expandafter\def\csname LT7\endcsname{\color{black}}%
      \expandafter\def\csname LT8\endcsname{\color{black}}%
    \fi
  \fi
  \setlength{\unitlength}{0.0500bp}%
  \begin{picture}(3744.00,2880.00)%
    \gplgaddtomacro\gplbacktext{%
      \csname LTb\endcsname%
      \put(528,550){\makebox(0,0)[r]{\strut{}\scriptsize 0}}%
      \put(528,728){\makebox(0,0)[r]{\strut{}}}%
      \put(528,906){\makebox(0,0)[r]{\strut{}}}%
      \put(528,1084){\makebox(0,0)[r]{\strut{}}}%
      \put(528,1262){\makebox(0,0)[r]{\strut{}}}%
      \put(528,1440){\makebox(0,0)[r]{\strut{}\scriptsize 1}}%
      \put(528,1617){\makebox(0,0)[r]{\strut{}}}%
      \put(528,1795){\makebox(0,0)[r]{\strut{}}}%
      \put(528,1973){\makebox(0,0)[r]{\strut{}}}%
      \put(528,2151){\makebox(0,0)[r]{\strut{}}}%
      \put(528,2329){\makebox(0,0)[r]{\strut{}\scriptsize 2}}%
      \put(660,330){\makebox(0,0){\strut{}\scriptsize 0}}%
      \put(942,330){\makebox(0,0){\strut{}}}%
      \put(1224,330){\makebox(0,0){\strut{}\scriptsize 0.2}}%
      \put(1506,330){\makebox(0,0){\strut{}}}%
      \put(1788,330){\makebox(0,0){\strut{}\scriptsize 0.4}}%
      \put(2070,330){\makebox(0,0){\strut{}}}%
      \put(2351,330){\makebox(0,0){\strut{}\scriptsize 0.6}}%
      \put(2633,330){\makebox(0,0){\strut{}}}%
      \put(2915,330){\makebox(0,0){\strut{}\scriptsize 0.8}}%
      \put(3197,330){\makebox(0,0){\strut{}}}%
      \put(3479,330){\makebox(0,0){\strut{}\scriptsize 1}}%
      \put(154,1439){\rotatebox{-270}{\makebox(0,0){\strut{}\small $\phi/\pi$}}}%
      \put(2069,110){\makebox(0,0){\strut{}\small $k/N$}}%
      \put(2069,2527){\makebox(0,0){\strut{}\small c)}}%
    }%
    \gplgaddtomacro\gplfronttext{%
    }%
    \gplbacktext
    \put(0,0){\includegraphics{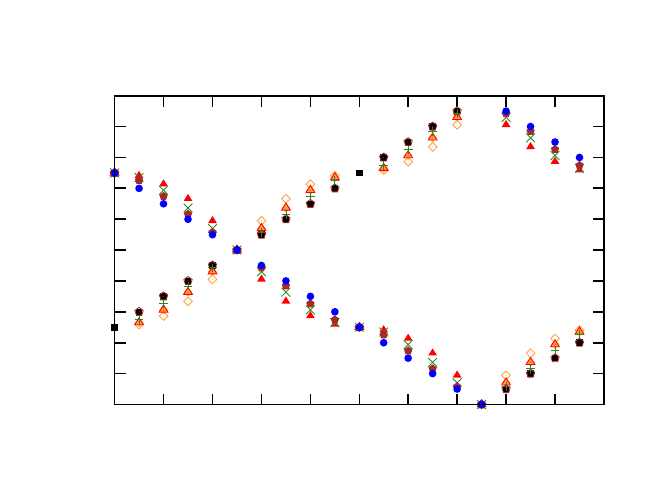}}%
    \gplfronttext
  \end{picture}%
\endgroup